\documentclass[preprint,12pt]{elsarticle}

\usepackage{graphics}
\usepackage{graphicx}
\usepackage{epsfig}

\usepackage[usenames]{color}

\usepackage{amssymb}

\biboptions{square, comma}

\begin{document}

\begin{frontmatter}

\title{
A new approach to microscopic modeling of a hole transfer in heteropolymer DNA
}

\author{A. S. Shigaev\corref{cor1}}
\ead{shials@rambler.ru}
\author{O. A. Ponomarev}
\ead{olegpon36@mail.ru}
\author{V. D. Lakhno\corref{cor2}}
\ead{lak@impb.psn.ru}

\cortext[cor1]{Corresponding author}
\cortext[cor2]{Principal Corresponding author}

\address{Institute of Mathematical Problems of Biology, Russian Academy of Sciences, 142290, Pushchino, Moscow Region, Russia}

\begin{abstract}
Thermal oscillations of base pairs in the Peyrard-Bishop-Holstein model are simulated by stochastic fluctuations of base overlap integrals. Numerical investigation of the model is carried out for a hole transfer in $G_1A_2G_3G_4$ sequence which was previously studied experimentally by F. Lewis et al. A hole migration between $G_1$ and $G_3G_4$ is determined by the matrix elements of the charge transition, but presence and amplitude of their stochastic fluctuations proved to play a key role in reproduction of the experimental kinetics. Good agreement with the experimental data was obtained for a wide range of the model parameters' combinations.
\end{abstract}


\end{frontmatter}

\section{Introduction}

Investigations of DNA conducting properties are very important for both classical radiobiology \cite{1, 2} and quite a new science of nanobioelectronics \cite{3, 4}. Charge migration in DNA is strongly dependent on a set of conditions, the basic of which is the nucleotide sequence. Researchers are generally focused on a cation-radical (hole) migration. A detailed consideration of experimental and theoretical material on charge transfer in DNA is presented, for example, in reviews by Conwell \cite{5} and Wagenknecht \cite{6}.

Experimental investigations of a hole transfer in DNA can be divided into steady-state and time-resolved measurements. The first ones are based on ionization of modified DNA followed by the analysis of its oxidation products by electrophoresis, HPLC or other methods. Owing to steady-state methods it has been found, that a cation-radical can move in DNA over a distance of tens or even hundreds of angstroms \cite{7} -- \cite{9}. Theoretical studies demonstrated that in the case of heterogeneous DNA chains, a hole transfer is realized as a series of hops between guanine nucleobases, which have the lowest oxidation potential \cite{10} -- \cite{12}. Theoreticians discuss two basic models of the cation-radical transition, which were originally suggested by J. Jortner et al. \cite{13}. In the first, superexchange model, a charge tunnels immediately from a donor to an acceptor, avoiding chemical interaction with intervening nucleotides. The second, multistep hopping model, implies a successive transition of a hole from one base to another due to the energy of thermal fluctuations. In the case of regular and homogeneous chains the transition can also be realized by a band, polaron, soliton, breather, etc. mechanisms \cite{14}.

Though the equilibrium methods have played a major role in the study of charge transfer, the information of relative hopping rates is incomplete for estimating the time of a hole migration over a particular distance. By contrast, time-resolved measurements enable one to get absolute rates of separate charge hopping steps in a DNA fragment \cite{15} -- \cite{18}.

Prominent among all time-resolved measurements are the works by F. Lewis et al. \cite{16, 17}. Unlike many other works (see, for example, \cite{15, 18}), there the object of investigation was oligonucleotides shorter than 10 base pairs. One end of these oligomers was bound to stilbene-4,4'-
dicarboxamide chromophore (hereafter St) which actually made them hairpins. A hole transfer in such structures takes place in nanosecond timescale. The general form of hairpins from work \cite{16} was
$St$-$A_m$-$G_1A_2G_3G_4$-$A_n$,
where
$1 \le m \le 3$, and $0 \le n \le 2$.
Hereafter, 1 -- 4 are numbers of bases in the "key fragment". The singlet stilbene selectively photooxidizes $G_1$, resulting in the formation of a primary radical-ion pair
 \{$St^{- \bullet}$ \ \ \ \ \  $G_1^{+ \bullet}A_2G_3G_4$\}.
In the subsequent reversible transfer of a hole to the $G_3G_4$ doublet $G_1^{+ \bullet}$ in turn acts as an electron acceptor. Hence, both a donor and an acceptor are guanine bases in the hairpins under study. This fact together with a small size and simple primary structure of the hairpins leads to a simple reaction scheme. The scheme enables direct assessment of the rate constants for a hole transfer from $G_1$ to $G_3G_4$ and back.

Due to a simple and elegant technique F. Lewis et al. were the first to determine in 2000 the values for rate constants of the forward and return cation-radical transport ($k_t$ and $k_{-t}$ respectively). Their values were 50 and 6 $\mu s^{-1}$ respectively. These data are in good agreement with other time-resolved measurements (see, for example, \cite{18}).

Nevertheless, the theoretical interpretation of the charge transfer dynamics presented in \cite{16} is based on a phenomenological description with the use of a reaction scheme. It does not make possible studying physical principles of the cation-radical migration in heteropolymer DNA. The aim of this work is to describe the results of \cite{16} by a microscopic model and explain them relying on the data obtained in the computational experiment.

Here we studied numerically the Peyrard-Bishop-Holstein model for a fragment of a $G_1A_2G_3G_4$ sequence. The model allows an accurate taking account of the base-pair dynamics in terms of the Peyrard-Bishop-Dauxois Hamiltonian, which has been sucessfully applied for the study of various aspects of DNA denaturation for over 20 years \cite{19} -- \cite{21}. In our approach thermal oscillations of the lattice structure are introduced via small stochastic fluctuations of nondiagonal matrix elements of the charge transition in the quantum subsystem.

These matrix elements present the only group of parameters that was obtained in quantum-chemical computations (see refs below). This fact is very important: taking into account the influence of the temperature, hydration and short length of the duplex, nondiagonal elements can possess some other values in real DNA in water solution \cite{22, 23}. Hereafter the term "parameter combination" implies combination of values of the nondiagonal matrix elements and amplitudes of their small stochastic fluctuations. All the other parameters of the Peyrard-Bishop-Holstein model are experimentally measured quantities and remained invariant in our simulations.

In the numerical experiment, a hole transfer over the $G_1A_2G_3G_4$ fragment is investigated for more than 100 parameter combinations. These simulations enabled us to find a set of parameter combinations at which the charge transfer rates averaged over the "microensemble" are in good agreement with the experimental data obtained by Lewis and co-workers. So, the experimental data on the cation-radical transfer along DNA are for the first time reproduced by microscopic modeling. Below we describe in detail the model and the approach developed, discuss our results and compare them with the experiment.

\section{The model}

The Peyrard-Bishop-Holstein (PBH) model is a recently developed "hybrid" of the Holstein \cite{24} and Peyrard-Bishop-Dauxois \cite{19} models. So far, the PBH model has been used to study the charge migration only in homopolymer DNA \cite{25} -- \cite{27}. In the studied case of a heterogeneous sequence, the motion equations for a hole and nucleotide pairs in the neighborhood approximation have the form:
\begin{eqnarray}
\nonumber
i\hbar\frac{d\psi_{n}}{d\tilde t} =
\nu_n^0\psi_{n} + \delta^\prime \tilde u_n \psi_n +
\nu_{n,n-1}\psi_{n-1} + \nu_{n,n+1}\psi_{n+1}
\\
m\frac{d^2\tilde u_n}{d\tilde t^2} = -
\frac{\partial}{\partial \tilde u_n}V(\tilde u_n) -
\frac{\partial}{\partial \tilde u_n}W(\tilde u_n, \tilde u_{n+1}) -
\\
\nonumber
\frac{\partial}{\partial \tilde u_n}W(\tilde u_{n-1}, \tilde u_n) -
\delta^\prime|\psi_n|^2    -    \gamma\frac{d\tilde u_n}{d\tilde t}
\end{eqnarray}
In the first equation, $\hbar$ is the Planck constant, $\psi_n$ --- the probability amplitude for the charge carrier located at the n-th site, $\nu_n^0$ --- its oxidation potential, $\delta^\prime$ is the charge-vibrational coupling constant, $\tilde u_n$ represents the transverse displacement of the hydrogen bonds connecting two bases, $\nu_{n,n \pm 1}$ are matrix elements of the transition.
  In the second equation, $m$ is the effective site mass, $\gamma$ --- friction constant and
\begin{eqnarray}
V(\tilde u_n) = \tilde D_n (e^{-\tilde \alpha_n\tilde u_n}  -  1)^2
\end{eqnarray}
\begin{eqnarray}
W(\tilde u_{n-1}, \tilde u_n) = \frac{k}{2}
(1 + \rho e^{-\tilde \beta(\tilde u_n + \tilde u_{n-1})})
(\tilde u_n - \tilde u_{n-1})^2
\end{eqnarray}
where $\tilde D_n$ and  $\tilde \alpha_n$ determine the depth and width of the energy well in Morse potential $V(\tilde u_n)$. In the nearest-neighbor potential $W(\tilde u_{n-1}, \tilde u_n)$, accounting for stacking interactions, $k$ --- coupling constant, $\tilde \beta$ --- damping coefficient and $\rho$ is dimensionless stiffness parameter. The condition $\rho \not= 0$ reproduces the fact, that the double-stranded backbone is more rigid than the unwound strands.

The friction constant was taken to be equal to $6 \cdot 10^{-13} N \cdot m^{-1} \cdot s$, which corresponds to the picosecond characteristic time scale of DNA oscillations \cite{28}.  The other parameters of the classical subsystem were taken from the well-known Letter by Campa and Giansanti:
$\tilde D_{AT}$ = 0.05 eV,
$\tilde D_{GC}$ = 0.075 eV,
$\tilde \alpha_{AT}$ = 4.2 \AA$^{-1}$,
$\tilde \alpha_{GC}$  = 6.9 \AA$^{-1}$,
$k$ = 0.025 eV $\cdot$ \AA$^{-2}$,
$\tilde \beta$ = 0.35 \AA$^{-1}$,
$\rho$ = 2 \cite{29}.
The oxidation potentials $\nu_n^0$ for adenine and guanine were taken from experimental data for acetonitrile solution: $\nu_A^0$ = 1.69 eV, $\nu_G^0$  = 1.24 eV \cite{30}. The potential for $G$, being the lowest one, was taken as zero and, hence, $\nu_A^0$  = 1.69 $-$ 1.24 = 0.45 eV. The estimated values of the matrix elements on the $G_1A_2G_3G_4$ fragment were also taken from literature: $\nu_{GA}$  = 0.089 eV, $\nu_{AG}$  = 0.049 eV and  $\nu_{GG}$ = 0.084 eV \cite{22}.

When transforming system (1) into dimensionless form, the condition was specified:
\begin{eqnarray}
\nonumber
\frac{\delta^\prime \tau^2}{m \cdot U} =
\delta^\prime \cdot U \frac{\tau}{\hbar} = \chi
\end{eqnarray}
where $\tau$ --- the timescale of the system, $U$ is an arbitrary scale of the transverse displacement, $\chi$ -- the dimensionless form of the charge-vibrational coupling constant. If we choose $\delta^\prime$ = 0.13 eV $\cdot$ \AA$^{-1}$ \cite{28}, then $U$ = $10^{-12}$ m and $\chi \approx 0.02$, provided that $m$ = $10^{-24}$ kg and $\tau$ = $10^{-14}$ s. Consequently, the parameters of the model relate to their dimensionless forms as
\begin{eqnarray}
\nonumber
\tilde \alpha_{n} = \alpha_{n} \cdot U^{-1};
\quad
\tilde t = t \cdot \tau;
\quad
\tilde \beta = \beta \cdot U^{-1};
\quad
\tilde u_n = u_n \cdot U;
\end{eqnarray}
\begin{eqnarray}
\nonumber
\tilde D_{n} = D_{n} \cdot m \cdot U^{2} \cdot \tau^{-2};
\quad
\gamma = \omega^\prime \cdot m \cdot \tau^{-1};
\quad
k = \Omega^{2} \cdot m \cdot \tau^{-2};
\end{eqnarray}
\begin{eqnarray}
\nonumber
\nu_n^0 = \eta_n^0 \cdot \hbar \cdot \tau^{-1};
\quad
\nu_{n,n \pm 1} = \eta_{n,n \pm 1} \cdot \hbar \cdot \tau^{-1};
\end{eqnarray}
where $\eta_{n,n \pm 1}$, $\eta_n^0$, $\omega^\prime$ and $\Omega^{2}$ --- dimensionless values of $\nu_{n,n \pm 1}$, $\nu_n^0$, $\gamma$ and $k$ respectively.

The dimensionless form of system (1) was realized in a parallel MPI-program.  Numerical calculations were carried out by the fourth-order Runge-Kutta method, the step size being 10$^{ -18}$ s. Each separate realization was simulated for over 30 ns (3 $\cdot$ 10$^{10}$ steps). At the initial moment the charge was on $G_1$, $\psi_1 = 1$. To specify boundary conditions we introduced fictitious terminal base pairs $X_0$ and $X_5$, for which $\nu_{n, n\pm 1}$  and $\tilde u_n$  had a fixed value --- zero. Hence, hole migration to $X_0$ and $X_5$ in the $X_0G_1A_2G_3G_4X_5$ sequence was impossible. Such boundary conditions were referred to as fixed.

\section{The method}

In a real duplex, thermal vibrations of the bases lead to fluctuations of overlap integrals of their $\pi$-orbitals. Since DNA is a complex molecule with a generous amount of vibrational modes, the time dependence of $\eta_{n, n\pm 1}$ is a very complicated stochastic function. Nevertheless, at this stage of modeling a key moment is taking account of fluctuations of non-diagonal matrix elements itself. Actually, a more careful consideration of these perturbations is desirable, but in the first approximation it is enough to specify the only (constant) frequency of $\eta_{n, n\pm 1}$ fluctuations. We take this parameter to be equal to the frequency of the stretched oscillations (along H-bonds), i. e. 0.5 $\cdot$ 10$^{12}$ s$^{-1}$. In this case, the amplitude of each individual oscillation is a random value, whereas the root-mean-square amplitude of the fluctuations is a parameter of the model.

In our numerical experiments thermal oscillations were simulated by random deviations of the elements $\eta_{GA}$, $\eta_{AG}$ and $\eta_{GG}$  from their "reference" values specified as parameters. The reference value of any matrix element (let us call it $\eta_b$ ) was invariable throughout the realization. Fluctuations were provided by the stochastic addition $\eta$. In such a way, the resulting value of the matrix element was the sum $\eta_b + \eta$. Generated pairs of random numbers $r_1$ and $r_2$ whose values were in the range from 0 to 1, were then subjected to Box-Muller transformation \cite{31}. This resulted in $\eta_a$ --- a random addition $\eta$ in the next point of extremum
\begin{eqnarray}
\nonumber
\eta_a = S \cdot cos (2 \pi r_1) \cdot \sqrt{-2 \ln r_2}
\end{eqnarray}
 where $S$ is the root-mean-square amplitude of a random deviation.

At initial moment $\eta$ was equal to zero. At that very moment $\eta_a$ was generated for the next fluctuation: $\eta_a = \eta(t + 10^{-12} s)$. In the course of the simulation $\eta$ changed with the step $j = 10^{-14}$ s according to the law
\begin{eqnarray}
\nonumber
\eta(t + j) = \frac{\eta_a^\prime + \eta_a}{2} + \frac{\eta_a^\prime - \eta_a}{2} \cdot
cos(\pi \cdot j \cdot 10^{12} s^{-1}),
\end{eqnarray}
where $\eta_a^\prime$ is the value of the previous deviation (initially equal to zero), and $j$ = $1 \cdot 10^{-14}$ s, $2 \cdot 10^{-14}$ s, \dots , $99 \cdot 10^{-14}$ s, $10^{-12}$ s. At the moment $j = 10^{-12}$ s $\eta_a^\prime$ reached a value of $\eta_a$, while $\eta_a$ was generated anew.

For any combination of the matrix elements, the values of their standard deviations $S_{GA}$, $S_{AG}$ and $S_{GG}$ always related to one another by the constant ratio of 7 : 5 : 8. This was caused by the ratio of the quantities $\eta_{GA}$, $\eta_{AG}$ and $\eta_{GG}$ per se (8.5 : 4.7 : 8). A slightly reduced value of $S_{GA}$  reflects the fact that $G_1$ resides in the center of the duplex and is subjected to strong fluctuations less than the other base pairs under consideration. For each combination of the parameters $S_{GA}$, $S_{AG}$, $S_{GG}$, $\eta_{GA}$, $\eta_{AG}$ and $\eta_{GG}$ we performed 80 realizations to obtain the curves $\biggl\langle|\psi_1|^2\biggr\rangle_{80} = f(t)$, where angular brackets denote averaging.

\section{Results and discussion}

According to the data by A. Voityuk et al. \cite{22}, $\eta_{GA} = 1.351$, $\eta_{AG} = 0.743$, and $\eta_{GG} = 1.275$. Taking into account hydration of oligonucleotide, we specified variation of these quantities to be approximately 25\%. Thus, the reference values ($\eta_b$, see above) of the matrix elements was: $\eta_{GA} = 1.051, 1.351, 1.651$; $\eta_{AG} = 0.543, 0.743, 0.943$; $\eta_{GG} = 1.275, 0.975$. We excluded variant $\eta_{GG} = 1.575$ because of so-called "end-fraying", causing substantial reduction of corresponding overlap integral \cite{32} -- \cite{34}. The maximal standard deviations of the nondiagonal matrix elements was chosen such, that the probability of a situation $\eta_b + \eta_a < 0$ be less than 10$^{-9}$.

In the first series of simulations we took maximum dispersion of the matrix elements of the charge transition: $S_{GA} = 0.14$, $S_{AG} = 0.10$, $S_{GG} = 0.16$. The transfer was studied during 14 ns. For $\eta_{GG} = 1.275$, in all the cases, most of the charge density remained on $G_1$ throughout the realization. The only exception was combinations of the parameters $\eta_{GA} = 1.651, \eta_{AG} = 0.743$ and $\eta_{GA} = 1.651, \eta_{AG} = 0.943$, where very slow transfer was observed. In the case of $\eta_{GG} = 1.025$ the charge escape from the first site took place for all the chosen $\eta_{GA}$ and $\eta_{AG}$ though its rate was quite low. We hypothesized that for any combination of $\eta_{GA}$ and $\eta_{AG}$ (in some range of these parameters) there exist a $\eta_{GG}$ and $S_{GA}$, $S_{AG}$, $S_{GG}$ such that the rates of direct and backward transfer of a hole over the $G_1A_2G_3G_4$ fragment are close to the experimental values obtained by Lewis et al, i.e. $k_t \approx 50 \mu s^{-1}$, $k_{-t} \approx 6 \mu s^{-1}$ \cite{16}.

In the second series of simulations we extended the time up to 30 ns and studied the model using 41 combinations of $\eta_{GA}$, $\eta_{AG}$ and $\eta_{GG}$ for strong fluctuations ($S_{GA} = 0.14$, $S_{AG} = 0.10$, $S_{GG} = 0.16$), using 54 combinations for moderate fluctuations ($S_{GA} = 0.07$, $S_{AG} = 0.05$, $S_{GG} = 0.08$) and using 27 combinations for weak fluctuations ($S_{GA} = 0.035$, $S_{AG} = 0.025$, $S_{GG} = 0.04$). The solution of kinetic equations, describing the process
\begin{eqnarray}
G_1^{+ \bullet}A_2G_3G_4 \leftrightarrows  G_1A_2G_3G_4^{+ \bullet}
\end{eqnarray}
which had the form
\begin{eqnarray}
|\psi_1|^2(t)_{approx} = F \frac{ k_{-t} +  k_t\cdot\exp[-(k_t + k_{-t})\cdot t)] }{k_t + k_{-t}}
\end{eqnarray}
was used to fit the curves $\biggl\langle|\psi_1|^2\biggr\rangle_{80} = f(t)$ obtained for each combination of the parameters. Constants $k_t$, $k_{-t}$ and the constant of proportionality $F$ were chosen such that the root-mean-square difference between $|\psi_1|^2(t)_{approx}$ and $\biggl\langle|\psi_1|^2\biggr\rangle_{80}(t)$ on the time interval from 10 to 30 ns be minimum.

For one of the functions $\biggl\langle|\psi_1|^2\biggr\rangle_{80}(t)$, together with finding the constants $k_t$ and $k_{-t}$, we carried out comparison with experimental data by Lewis. In the kinetic scheme of Letter \cite{16} consideration is given to two states of an ionized hairpin, namely, a radical-ion pair $X$ having the form of
\{$St^{- \bullet}$ \ \ \ \ \  $G_1^{+ \bullet}A_2G_3G_4$\}
and capable of recombination and a radical-ion pair $Y$
\{$St^{- \bullet}$ \ \ \ \ \  $G_1A_2G_3G_4^{+ \bullet}$\},
which cannot recombine. In our computational experiment a recombination reaction is lacking, therefore a straightforward comparison of our results with the experimental data by Lewis et al. is impossible. Nevertheless, knowing $ [X](t) \bigg/ ({[X](t) + Y(t)}) = \biggl\langle|\psi_1|^2\biggr\rangle_{80}(t)$ at every instant, we can carry out "indirect" substitution of our data into the kinetic scheme from Letter \cite{16}. Comparison of the computational data with the experiment is presented in Fig. 1.

\begin{figure*}
\includegraphics[width=0.9\textwidth]{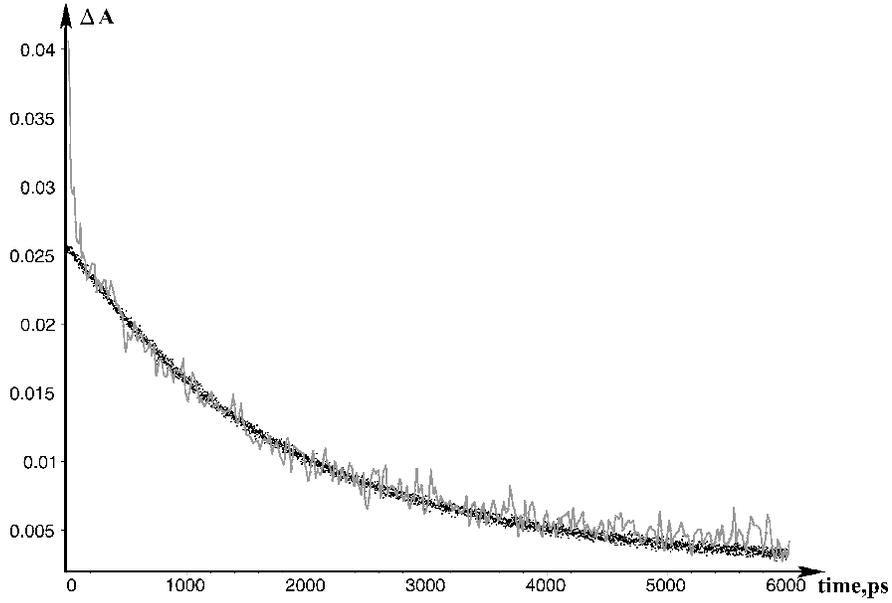}
\caption{Comparison of computational data obtained for the parameter set $S_{GA} = 0.07$, $S_{AG} = 0.05$, $S_{GG} = 0.08$, $\eta_{GA} = 1.051$, $\eta_{AG} = 0.543$, $\eta_{GG} = 0.519$, with the experimental data found in \cite{16}. The estimated values of the constants: $k_t = 60 \mu s^{-1}$, $k_{-t} = 7.8 \mu s^{-1}$ \cite{16}. The best fitting to Lewis' data was achieved when $k_{cr} = 470 \mu s^{-1}$, the quantum yield and recombination being taken into account too. The solid line (gray) represents the experimental decay curve of $St^{- \bullet}$ from \cite{16}, where $\delta A$ is a relative transient absorption at 575 nm. The data of the computational experiment are presented as the dot set (black), $\biggl[ t , \ \ \  \biggl\langle|\psi_1|^2\biggr\rangle_{80}(t)\biggr]$. Time interval between the dots is equal to 2 ps.}
\label{fig_1}
\end{figure*}

For almost all combinations of  $\eta_{GA}$, $\eta_{AG}$ and $\eta_{GG}$, the closest to Lewis' values of $k_t$, $k_{-t}$ and equilibrium constants  were obtained for intermediate standard deviations: $S_{GA} = 0.07$, $S_{AG} = 0.05$ and $S_{GG} = 0.08$. These data are presented in Table 1. It turns out that $S$ plays a key role in the relation between $k_t$ and $k_{-t}$, obtained in the calculations: strong fluctuations of the matrix elements lead to a decrease in $k_t$ and increase in $k_{-t}$, that is to a decrease in the equilibrium constant. For example, the combination of the parameters $S_{GA} = 0.14$, $S_{AG} = 0.10$, $S_{GG} = 0.16$, $\eta_{GA} = 1.351$, $\eta_{AG} = 0.743$, $\eta_{GG} = 0.705$, yields $k_t = 29 \mu s^{-1}$, $k_{-t} = 20 \mu s^{-1}$: the equilibrium constant turns out to be almost 6 times less than the experimental value. Similarly, too weak fluctuations lead to abnormally large (more than 20) equilibrium constants due to negligible $k_{-t}$: 0.01 -- 0.5 $\mu s^{-1}$. Moreover, a standard deviation determines the dynamics of $\biggl\langle|\psi_1|^2\biggr\rangle_{80}(t)$ decay during the first few nanoseconds. At large $S$ the initial escape of a hole from $G_1$ occurs very fast ($k_t > 100 \mu s^{-1}$ during the first 3 - 4 ns). In the case of too weak fluctuations, by contrast, the initial rate of a hole migration is rather low as compared to its subsequent transfer.

\begin{table}
\caption{The values of rate constants $k_t$ and $k_{-t}$, obtained for various combinations of $\eta_{GA}$, $\eta_{AG}$ and $\eta_{GG}$. For each combination of $\eta_{GA}$, $\eta_{AG}$, the value of $\eta_{GG}$ is presented such that $k_t$ and $k_{-t}$  (which are also shown) are closest to the experimental data \cite{16}.
$\ \ \ \ \ \ \ $ $^a$ For the $\eta_{GA} = 1.651$, $\eta_{AG} = 0.943$, $\eta_{GG} = 0.9$, a great deviation of $k_{-t}$ from the experimental one results from insufficiently high amplitude of $\eta$ fluctuations: for $S_{GA} = 0.14$, $S_{AG} = 0.10$ and $S_{GG} = 0.16$ the values of $k_t$ and $k_{-t}$ are equal to 32 and 8.6 $\mu s^{-1}$, respectively, which is slightly closer to the data by Lewis et al \cite{16}.}
\begin{center}
\begin{tabular}[h]{| p{3em} | p{8em} | p{8em} | p{8em} |}
\hline
$\eta_{AG}$  &  0.543  &  0.743  &  0.943  \\ \cline{1-1}
$\eta_{GA}$ &  & & \\
\hline
1.051  &  $\eta_{GG} = 0.52 \pm 0.02$

$k_t = 59.2 \mu s^{-1}$

$k_{-t} = 7.2 \mu s^{-1}$       &
$\eta_{GG} = 0.56 \pm 0.02$

$k_t = 53.9 \mu s^{-1}$

$k_{-t} = 6 \mu s^{-1}$    &
$\eta_{GG} = 0.56 \pm 0.02$

$k_t = 67.5 \mu s^{-1}$

$k_{-t} = 7 \mu s^{-1}$     \\
\hline
1.351    &   $\eta_{GG} = 0.675 \pm 0.015$

$k_t = 52.2 \mu s^{-1}$

$k_{-t} = 6 \mu s^{-1}$       &
$\eta_{GG} = 0.71 \pm 0.02$

$k_t = 54.5 \mu s^{-1}$

$k_{-t} = 5.5 \mu s^{-1}$    &
$\eta_{GG} = 0.73 \pm 0.02$

$k_t = 48.3 \mu s^{-1}$

$k_{-t} = 7.8 \mu s^{-1}$    \\
\hline
1.651  &   $\eta_{GG} = 0.825 \pm 0.02$

$k_t = 52.5 \mu s^{-1}$

$k_{-t} = 5.85 \mu s^{-1}$       &
$\eta_{GG} = 0.87 \pm 0.02$

$k_t = 46.4 \mu s^{-1}$

$k_{-t} = 4 \mu s^{-1}$    &
$^a$ ${\eta_{GG}} = 0.9 \pm 0.03$

$k_t = 42 \mu s^{-1}$

$k_{-t} = 0.15 \mu s^{-1}$    \\
\hline
\end{tabular}
\end{center}
\end{table}

The data of Table 1 suggest that one cannot unambiguously determine the values of the matrix elements of the charge transition at which $k_t$ and $k_{-t}$ fits the experimental data. Moreover, the range of $\eta_{GA}$, $\eta_{AG}$, and $\eta_{GG}$ combinations for which the rate of the charge migration corresponds to the experiment is, probably, rather wide.

It may seem strange that all the values of $\eta_{GG}$  in Table 1 are 1.5 -- 2.5 times less than the theoretical one (1.275 dimensionless units). However this deviation can easily be explained by end effects. In this respect very illustrative are NMR-investigations of local "flipping-out" of bases in DNA oligomers. Due to the end-fraying, the rates of flipping of terminal base pairs are so high that they cannot be measured \cite{32} -- \cite{34}. Besides, the influence of the end-fraying can extend to the second and even the third base pair \cite{32, 33}. In experiments by Lewis et al. end-fraying could be responsible for reduced overlapping of two terminal base pairs. On the other hand, deviation of $\eta_{GG}$  from the literature value may be to some extent caused by imperfection of our calculation technique. Supposedly, if we change the fixed boundary conditions, for example, to the periodical ones, we will manage to get the values of $\eta_{GG}$  more close to the estimated data by Voityuk et al. \cite{22}.

In this Letter we studied migration of a cation-radical in the $G_1A_2G_3G_4$ fragment in the framework of a proposed semistochastic variant of the PBH model. We have succeeded to show the existence of a wide range of the model parameter combinations at which the rates of direct and reverse charge transfer are close to the experimental data. In the studies of a hole transfer in the framework of a more simple microscopic model, the time averaged rate of a hole transfer from $G_1$ to $G_3G_4$ was approximately two orders of magnitude higher than in the experiment \cite{35}. Moreover, the shape of the curve $|\psi_1|^2 = f(t)$ per se reminded exponential decay only slightly. In our computational results, only the function $\biggl\langle|\psi_1|^2\biggr\rangle_{80} = f(t)$  has an exponential form, while $|\psi_1|^2 = f(t)$ for each individual realization has nothing to do with an exponent. Due to features of Morse potential (see eq. (2)) a positive charge cannot substantially deform the molecular lattice (in this case $\tilde u_n < 0$). This dramatically constraints its mobility. Therefore, the "motor" of a hole migration is transient and rather rare "successful" combinations of the matrix elements of the charge transition.

The problem of how the transfer of a cation-radical relates to fluctuations of base overlap integrals is to be studied in the future. Further development of our approach implies the direct coupling of matrix elements' fluctuations with thermal dynamics of DNA in PBH model (Langevin approach, see, e. g., \cite{36}). An alternative way is the design of a more quick semistochastic algorithm of charge transport simulation in a long heterogeneous DNA. For example, we intend to model a hole transfer in other sequences studied by Lewis et al. \cite{17}.

Microscopic modeling of DNA has played a great role in the studies of DNA thermal and mechanical denaturation \cite{19} -- \cite{21}. In the same way it is promising for investigation of charge migration. For example, in thermodynamical calculations which are usually used for interpretation of experiments on a hole transfer in DNA, an appropriate taking account of chaotic fluctuations of overlap integrals is hardly feasible. Nevertheless, an important role of these fluctuations in providing the charge transfer along heteropolymer DNA was proved experimentally \cite{37}.

Here a computational experiment enabled us to demonstrate an important role of stochastic oscillations of the bases in the cation-radical migration process. Moreover, consideration of these oscillations turned out to be of crucial importance for reproduction of experimental data on charge transfer in heteropolymer DNA.

\section{Acknowledgements}

The work was done with the support from the RFBR, projects 10-07-00112-a; 11-07-00635-a. The authors are thankful to the Joint Supercomputer Center of the Russian Academy of Sciences, Moscow, Russia for the provided computational resources.

\bibliographystyle{model1a-num-names}

\end{document}